# Information flow simulation community detection of weighted-directed campus friendship network in continuous time

Ren Chao, Yang Menghui*

（School of Information Resource Management, Renmin University of China, Beijing 100872, China）

**Abstract:** Educational data mining has become an important research field in studying the social behavior of college students using massive data. However, traditional campus friendship network and their community detection algorithms, which lack time characteristics, have their limitations. This paper proposes a new approach to address these limitations by reconstructing the campus friendship network into weighted directed networks in continuous time, improving the effectiveness of traditional campus friendship network and the accuracy of community detection results. To achieve this, a new weighted directed community detection algorithm for campus friendship network in continuous time is proposed, and it is used to study the community detection of a university student. The results show that the weighted directed friendship network reconstructed in this paper can reveal the real friend relationships better than the initial undirected unauthorized friendship network. Furthermore, the community detection algorithm proposed in this paper obtains better community detection effects. After community detection, students in the same community exhibit similarities in consumption level, eating habits, and behavior regularity. This paper enriches the theoretical research of complex friendship network considering the characteristics of time, and also provides objective scientific guidance for the management of college students.

**Keyword**：campus friendship network, community detection, weighted directed network, campus behavior analysis.

---

* Electronic mail: renchao@ruc.edu.cn, yangmenghui@ruc.edu.cn

# 1.Introduction

In recent years, the development of data-driven science and social computing has sparked a surge of interest among scholars in simulating and studying human social behavior using massive data [1-2]. The evolution of the campus friendship network over time reflects the real-time changing social behavior of college students. Therefore, studying the evolution of the campus friendship network can provide us with valuable insights into the psychological and life states of students, which can help universities and teachers to provide better support [3]. Although some studies have achieved impressive results by constructing college students' campus friendship network using small-scale data obtained through questionnaire surveys and interviews, this method of data acquisition is subjective. Moreover, small-scale data is insufficient to accurately represent the real and objective campus friendship network [4]. Hence, it is crucial to study the campus friendship network using large-scale objective data recorded by the campus card system of universities [5].

Most of the previous research on campus friendship network focused on constructing a weighted-undirected graph based on students' consumption data. The co-occurrence of two students spending money in the same place within a short period is considered as an indicator of social bond strength, with more co-occurrence indicating stronger bonds [6]. However, this approach has two limitations. Firstly, the weighted-undirected graph cannot fully depict the real campus friendship network, which is always directed in the real world, since relationships between two people are usually unequal. Secondly, the invariable weight cannot show the dynamic evolution of the campus friendship network. To overcome these limitations, we reconstructed the weighted-undirected graph into a weighted-directed graph with tie decay in continuous time. Specifically, we introduced the concept of fitness, which represents the social ability of nodes in the campus friendship network [7]. According to the Bianconi-Barabasi model, node fitness distribution is equivalent to the power exponential distribution of node degree, which is linearly dependent on node degree. Thus, the greater the fitness or degree of the nodes in the campus friendship network, the stronger the social abilities of the students represented by this node, and the more friends and complex social relationships the student has. What's more, the weight between nodes in the campus friendship network changes over time, decaying exponentially. When co-occurrence occurs between nodes, the corresponding weight is immediately increased; between one co-occurrence and the next co-occurrence, the weight will show the trend of exponential decay [8]. Therefore, the direction of the campus friendship network in this paper is indicated by the nodes with high degree to the nodes with low degree. We reconstructed the weight of the campus friendship network based on the theory of tie decay in continuous time. The reconstructed weighted-directed graph with tie decay in continuous time accurately represents the real campus friendship network. In addition, social interactions in friendship network can form multiple communities, and the campus friendship network is no exception. Students in the same community tend to have more interactions, while those in different communities tend to have fewer interactions [9]. Therefore, community detection is an intuitive way to identify different friendship communities within the campus friendship network and understand how students' friendships change over time [10]. However, carrying out community detection on friendship network with large-scale dynamic data is a challenging task [11]. To address this challenge, this paper introduces a community detection algorithm based on Information Flow Simulation (IFS). We identify influential students in the campus friendship network who act as points of origin for information based on the reconstructed weighted-directed graph. Unlike the traditional IFS algorithm, we use the PageRank score of the weighted-directed graph in continuous time as the criterion for evaluating the influence of nodes. This allows us to detect communities in the evolving process of the campus friendship network in continuous time [12].

In this paper, we analyzed the co-occurrence patterns among students using student card consumption data from a university over the period 2017 to 2021, and constructed a campus friendship network based on these co-occurrence relationships. Building on the limitations of previous studies, we developed a new friendship network model based on a weighted-directed graph in continuous time that captures the temporal dynamics of student interactions more accurately. We then calculated the PageRank scores of nodes in the reconstructed network and used an Information Flow Strength (IFS) metric to identify the origin of information flow in the network. Higher PageRank scores indicate greater likelihood of being an origin of information flow, with information passing from the origin node to other nodes along the same path, defining a community. Our community detection analysis revealed that students in the same community exhibit significant homogeneity, both in academic performance and in employment choices after graduation. This finding has important implications for schools and teachers in predicting student behavior and offering appropriate support.

The paper is structured as follows: Section 2 provides an overview of the existing literature on campus friendship network and community detection. In Section 3, we describe our method for reconstructing the campus friendship network and conducting community detection. The results and analysis are presented in Section 4. Finally, in Section 5, we conclude the paper with a summary of our key findings and suggestions for future research.

## 2.Related work

### 2.1 Campus friendship network

The campus friendship network is a reflection of the friendship network that students form through their communication and behavior on campus. Studying this network can provide valuable insights into students' mental state and behavior patterns. Currently, there are many related studies in this field. For instance, Shyh-Chyang Lin et al. used friendship network analysis and psychological testing to investigate the friendship characteristics and learning habits of engineering majors in universities and established their friendship and learning network [13]. Other studies have focused on identifying the selection effect and socialization process underlying college student friendships, where a student's centrality in their peer network has been found to be linked to their academic performance [14]. Furthermore, it has been observed that a network of low-achieving friends can negatively impact a student's personal achievement, and an individual's behavior can also influence the formation of the network structure [15]. These studies highlight the significant impact of the campus friendship network on students' academic performance and personal behavior. However, previous research collected data through interviews or questionnaires, which may raise concerns about the data's objectivity and authenticity.

To ensure greater objectivity, some scholars have turned to student card data to study the friendship network of students on campus. Yongli Li et al., for instance, detected friendship network using large-scale data and demonstrated how the structural hole of academic friends and the homogeneity of friendship affect students' GPA ranking [16]. To address the time window size error, a sliding time window method was proposed to estimate the time and space co-occurrence times of students from student card data [3]. Huaxiu Yao et al. constructed students' social relations based on their campus behaviors and used a semi-supervised algorithm to predict their academic performance using the constructed friendship network [17]. To study the evolutionary characteristics of student friendship and the relationship between behavioral characteristics and student interaction, Zongkai Yang constructed four different evolutionary friendship network based on students' dynamic behavior data and found their trends [6]. Despite these accomplishments, there are still some limitations. The aforementioned campus friendship network are weighted-undirected, which overlooks the intrinsic properties of students in friendship network, such as the influence of individual communicative ability on friendship network. Furthermore, the temporal evolution of friendship is based on discrete time, and the choice of time window size can significantly affect the results. Therefore, it is essential to investigate the weighted-directed campus friendship network in conjunction with individual communicative ability.

### 2.2 Community detection

Community detection is a fundamental task in network analysis, aiming to identify tightly connected groups of vertices in the network. In the context of campus friendship network, community detection can reveal students' circles of friends and their impact on individual behavior. Various community detection algorithms have been developed, such as those based on deep neural networks, graph embeddings, and graph neural networks, which have achieved high accuracy in standard datasets [19]. Temporal community detection in evolving networks has also received considerable attention. For instance, Jialin He et al. proposed a fast algorithm for dynamic community detection in temporal networks that leverages information from the previous time step to improve efficiency while maintaining quality [18]. To identify stable communities in temporal networks, Hongchao Qin et al. developed a community detection algorithm based on a density map clustering framework [20]. Although these methods have yielded promising results, none can address the challenge of detecting communities in a weighted-directed campus friendship network in continuous time [12]. Therefore, this paper proposes a novel approach that leverages the PageRank score of nodes in the weighted-directed network to determine the origin of information flow, which is then used to accurately detect communities. By combining this approach with the traditional IFS method, we can achieve precise community detection in the directed campus friendship network in continuous time.

## 3. Methods

### 3.1 The construction of initial campus friendship network

In the campus friendship network, social relationships between students are determined by co-occurrence. This means that the more times two students were in the same place within a short period of time, the closer their social bond. In this study, we define two students to have co-occurred if they consumed at the same place within a 2-minute timeframe. Using this definition, we constructed an initial campus friendship network, where the nodes represent students and the weight of each edge represents the number of co-occurrences between two students. Here are the specific details.

There are $S$ students and $L$ locations, $t_m^l$ indicates that student $m$ has a consumption behavior at place $l$ at time $t$, $m, n = 1, 2, 3, \dots S$, $l = 1, 2, 3, \dots, L$, the calculation of the co-occurrence $w_{m,n}^l$ of student $m$ and $n$ at the location $l$ is as follows:

$$w_{m,n}^l = \begin{cases} 1, & if\ |t_m^l - t_n^l| \leq \Delta t \\ 0, & otherwise \end{cases} \quad (1)$$

where $\Delta t$ is time window, in this paper, we set $\Delta t = 120$ seconds.

The initial campus friendship network was built based on co-occurrence, in which each student is regarded as a node and the co-occurrence between two students is regarded as the weight. Weight between student $m$ and $n$ is $w_{m,n}$,

$$w_{m,n} = \sum_{l=1}^{L} w_{m,n}^{l} \qquad (2)$$

The initial campus friendship network is a weighted undirected graph, which can be expressed as a list of triples $(m, n, w_{m,n})$, where $m$ and $n$ represent the start and end points of an edge, respectively, and $w_{m,n}$ represents its weight.

## 3.2 The reconstruction of campus friendship network

### 3.2.1 Direction

Social skills play a crucial role in determining the structure of friendship network, with individuals who possess better social skills generally having more friends and occupying central positions within the network. However, existing studies on large-scale data building friendship network have largely ignored the impact of individual social skills on friendship network formation, including within the context of campus friendship network. In this study, we aim to address this limitation by incorporating students' social skills into our analysis.

To make students' social skills quantifiable, we introduce the concept of fitness. In the campus friendship network, node fitness represents a student's ability to convert casual relationships into long-term social relationships, which is a reflection of their social skills. Following the Bianconi-Barabassi model, fitness is an inherent attribute of nodes that is genetically determined. This framework allows us to determine the fitness of each node and its distribution function. Specifically, the distribution of node fitness follows a power exponential distribution of node degree, which is linearly dependent on node degree. In the campus friendship network, the degree of a node refers to the number of edges connected and the number of nodes connected.

To account for individual social skills in the construction of the campus friendship network, we assume that students with larger degrees in the network possess stronger social skills and are more likely to dominate social interactions with students with smaller degrees. Therefore, in the campus friendship network, the direction of edges should be from nodes with larger degrees to nodes with smaller degrees.

Specifically, for the initial campus friendship network represented by a triplet list $(m, n, w_{m,n})$, if node $i$ in the network connected with $p$ nodes, the degree of node $i$ equals $p$.

Suppose there is a node $i$ adjacent to the node $j$,

(1) If $k_i > k_j$, the triplet list is rewritten as $(i, j, w_{i,j})$, which indicates the node $i$ is the starting node and the node $j$ is the ending node of the edge $(i, j)$;

(2) If $k_i < k_j$, the triplet list is rewritten as $(j, i, w_{i,j})$, which indicates the node $i$ is the ending node and the node $j$ is the starting node of the edge $(i, j)$;

(3) If $k_i = k_j$, the triplet list is rewritten as $(i, j, w_{i,j})$ and $(j, i, w_{j,i})$, which indicates the node $i$ and the node $j$ are the starting node and the ending node of the edge $(i, j)$ for each other.

In this way, the campus friendship network becomes a weighted-directed graph, and the direction can simulate the influence of students' social skills on the campus friendship network structure.

### 3.2.2 Weight

In the process of network evolution, the strength of connections between two nodes in the network varies over time, often in continuous time rather than discrete time. To capture the characteristics of students' relationships that evolve over time in the campus friendship network, this paper introduces the tie-decay model to reconstruct the weight in the campus friendship network, resulting in a continuous time campus friendship network.

Specifically, this paper proposes that the weight between two students increases by 1 immediately after a co-occurrence, and then decays exponentially until the next co-occurrence. To validate this viewpoint, we provide its mathematical formulation.

For the initial campus friendship network represented by a triplet list $(m, n, w_{m,n})$, on the basis of the initial weight $w_{m,n}$, the reconstructed weight $w'_{m,n}$ can be calculated as

$$w'_{m,n} = -\alpha w_{m,n} + \tilde{a}_{m,n} \tag{3}$$

If student $m$ and student $n$ co-occur in time $\tau_{mn}^{(1)}, \tau_{mn}^{(2)}, \ldots$, then

$$\tilde{a}_{m,n}(t) = \sum_k \delta(t - \tau_{mn}^{(k)}) e^{-\alpha(t - \tau_{mn}^{(k)})} \tag{4}$$

where we represent an instantaneous interaction at $t = \tau$ as a pulse with the Dirac $\delta$-function. $k$ indicates the initial time and $t$ indicates the current time. For example, when $k = 20210101$, $t = 20210201$, we can obtain the change of weight in campus friendship network in continuous time from January 1, 2021 to February 1, 2021 by Eqs. (3) and (4).

In the above equations, $\alpha$ is the tie-decay parameter, which eliminates the error caused by different time window sizes. When choosing a value for α, it is perhaps intuitive to think about the half-life $\eta_{1/2}$ of a tie, as it gives the amount of time for a tie to lose half of its strength in the absence of new interactions. Given $\alpha > 0$, the half-life of a tie is $\eta_{1/2} = \alpha^{-1} \ln 2$. This paper referred to the reference [10] for the choice of parameter $\alpha$.

If there are initial conditions $w_{m,n}(0) = 0$, the solution of Eq (2) is $w_{m,n}(t) = H(t - \tau)e^{-\alpha(t-\tau)}$, $H(t)$ is the Heaviside step function. For further explanation of the tie-decay model, please refer to [10].

### 3.3 Community detection

The weighted-directed network constructed in this paper can be considered a dynamic temporal network, which presents many challenges for community detection. Accurately capturing the temporal characteristics of continuous time and their important influence on network evolution is a difficult task. However, this paper proposes that communities in the campus friendship network are often formed around influential nodes, which spread information to other nodes and bring them together into a community. These influential nodes can be seen as the origin of information flow, and any node that is part of the information flow is considered to belong to the same community. For example, for the origin $m$, if the information from the origin $m$ flows to the nodes $n, p, q$, we can consider that the nodes $n, p, q$ and origin $m$ belong to the same community. Note that if information from origin $m$ does not flow to any of the other nodes, origin $m$ does not belong to any community. That's the idea of community detection by Information Flow Simulation.

The community detection process in our study is challenging due to the dynamic temporal nature of the constructed weighted-directed network. To overcome this challenge, we propose a method that considers influential nodes as the origin of information flow in the campus friendship network. Specifically, we first calculate the PageRank score of each node in the continuous time network, where nodes with higher scores are assumed to be more influential. Then, we simulate information flow from each student as the origin, and identify all nodes that information passes through as belonging to the same community. The community detection process continues until the information flows through the entire network or the algorithm reaches a stable state. By applying this approach, we obtain the final community detection results.

The $n \times 1$ vector $\pi$ of PageRank scores, with $\pi > 0$ and $\|\pi\|_1 = 1$, is the leading-eigenvector solution of the eigenvalue problem

$$G^T \pi = \pi \tag{5}$$

where $G$ is the $n \times n$ rate matrix of a teleporting random walk:

$$G = \lambda(D^+W + cv^T) + (1 - \lambda)\mathbb{I}v^T = \lambda P + (1 - \lambda)\mathbb{I}v^T \tag{6}$$

where $P = D^+W + cv^T$, $W$ is the set of weight in the network. The matrix $D$ is the diagonal matrix of weighted out-degrees, and $D^+$ is its Moore–Penrose pseudo-inverse. $\mathbb{I}$ is the $n \times 1$ vector of 1s. $c$ is also a $n \times 1$ vector, when node $m$ with 0 out-degree, $c_m = 1$, and $c_m = 0$ otherwise. The $n \times 1$ distribution vector $v$ encodes the probabilities of each node to receive a teleported walker, $v_m = 1/S$ for all $m$.

In the weighted directed network in continuous time, the PageRank score $\pi(t)$ is

$$\pi^{k+1}(t) = \lambda P^T(t)\pi^k(t) + (1 - \lambda)v \tag{7}$$

where $k$ indicates the initial time and $t$ indicates the current time. $\lambda$ is the probability of a random walk in the PageRank algorithm, $\lambda \in (0, 1)$. In this paper, $\lambda = 0.85$. $P^T$ is the temporal transition matrix, which is calculated by

$$P(t) = D^+(t)W(t) + c(t)v^T \tag{8}$$

where $D(t)$ is the diagonal matrix of weighted out-degrees (i.e., the row sums of $B(t)$) at time $t$. Please refer to Ref. [12] for more details.

After obtaining the PageRank score for each node, we rank each node by PageRank score. The nodes in the top percentiles of $\varepsilon$ are taken as the origin of information flow. Each origin is assigned a community label, and information flows to other nodes with a certain probability.

In this paper, the probability of node $m$ propagating information to node $n$ is

$$P(m,n) = \left(\frac{w_{m,n}}{\delta^*(m)}\right)^\beta; \quad \beta \in (0,1) \tag{9}$$

where $w_{m,n}$ is the weight of edge between node $m$ and node $n$. $\delta^*(m)$ is the out-degree of node $m$, which is defined as the sum of the weights of all out-degree of that node:

$$\delta^*(m) = \sum_{x=1}^{S} w_{mx} \tag{10}$$

According to Ref. [12], we set $\beta = 1/4$. Then,

$$P(m,n) = \frac{\sqrt[4]{w_{m,n}}}{\sqrt[4]{\delta^*(m)}} \tag{11}$$

There are $S$ nodes in the campus friendship network of this paper, we select $X$ origin of the first $\varepsilon\%$, $X = \varepsilon S$. For the origins $x_i = 1, 2, \ldots, X$, and non-origins $y_j = 1, 2, \ldots, S - X$, we assigned the label $i$ for $x_i$. Information in $x_i$ flows to $y_j$ with probability $P(x_i, y_j)$. If the information in $x_i$ flows through $y_j$, the community label $i$ of $x_i$ is propagated to $y_j$, then both $x_i$ and $y_j$ belong to community $i$. If the information in $x_i$ does not flow through either $y_j$, then $x_i$ is considered isolated and does not belong to any community.

It should be noted that in this paper, we made the assumption that the information from the origin can only flow to non-origin nodes, and not to nodes that are also origins. Additionally, nodes that have already been labeled with a community cannot be labeled with another community. The community detection process is considered complete when all non-origin nodes have been labeled or when the algorithm reaches convergence.

### 3.4 The detailed steps of community detection

In this paper, the detail steps of community detection in continuous time weighted-directed campus friendship network are as follows.

Step 1: Calculated the co-occurrence between student $i$ and student $j$ according to Eqs. (1) and (2). An initial weighted-undirected campus friendship network was constructed with co-occurrence as the weight and each student as the node.

Step 2: Then calculated the degree of each node in the weighted-undirected campus friendship network. The weighted-directed campus friendship network was constructed by adding direction to the network on the basis of the nodes with large degree pointing to the nodes with small degree.

Step 3: According to Eqs. (3) and (4), the weight of weighted-directed campus friendship network was reconstructed to accurately describe the evolution of the network in continuous time.

Step 4: The static network at any time $t$ was captured from the weighted-directed campus friendship network under continuous time. In this network, we calculated the PageRank score $\pi(t)$ of nodes according to Eq. (7).

Step 5: $X$ nodes with a large Pagerank score are selected as the origin. Assuming that information flows from the origin to other nodes with a certain probability, the probability calculation has shown in Eqs. (9) and (11).

Step 6: If the information flow of the origin passes through some nodes, these nodes and the origin are considered to belong to the same community. When all non-origins are given community labels or the algorithm converges, the community detection was completed.

## 4. Experiments and analysis

### 4.1 Data description

The data used in this paper comprises records of student spending via student cards from 2017 to 2021 at a university in China. The spending data includes desensitized student IDs, time, place, and type of spending (recharge or spending) for over 27,000 students. Due to the impact of COVID-19, there is no data on consumption during the half-year when students attended classes at home in the 2019-2020 school year, and data for the 2020-2021 academic year is frozen at the end of March 2021. In the data preprocessing phase, all records whose consumption type is recharge were removed. Additionally, 33 restaurants, 6 bathrooms, 6 boiler rooms, and 4 stores were selected based on their locations for the study.

Ensuring data privacy is a top priority in this paper. All student information is processed anonymously, and student IDs in the original data are represented as encrypted strings. To ensure data security, all data is stored on the computer in the Information and Technology Center of an anonymous Chinese university. Data processing and analysis can only be performed on this computer, which cannot transmit data. The study and the use of the data were approved by the university.

The experiments in this paper are conducted on a semester-by-semester basis, covering the fall 2017 semester through the fall 2020 semester. The school was closed during the spring 2020 semester due to COVID-19, and hence no student data was collected during that semester. In total, six semesters were selected: the fall semester of 2017, the spring and fall semesters of 2018, the spring and fall semesters of 2019, and the fall semester of 2020. For each semester, a weighted directed campus friendship network under continuous time was reconstructed, and associations were tested for these six campus friendship network.

## 4.2 Comparative analysis

In order to evaluate the accuracy of the community detection algorithm proposed in this paper, the fall semester of 2019 data was used to test different algorithms on both an undirected unweighted friendship network and the reconstructed weighted directed friendship network. Since it is difficult to determine the true community labels of real-world networks, the modularity index was used as a measure of the strength of the network community structure. Modularity is a commonly used metric with a range of values from -0.5 to 1, with larger values indicating better community detection results. The Louvain and IFS community detection algorithms were used for the unweighted undirected friendship network, while the LP and TLW community detection algorithms were used in addition to the proposed algorithm for the reconstructed weighted directed friendship network. The modularity and average number of communities were used as evaluation metrics, and the results are presented in Table 1. The proposed algorithm in this paper utilized 1000 time points to calculate relationship weights for each semester, and the top 20% of nodes were selected as the origin of information dissemination in the community detection process.

The Louvain algorithm is a modularity-based community detection algorithm. It works by having nodes traverse the community labels of their neighbors and selecting the community label that maximizes the modularity increment. Once the modularity has been maximized, each community is treated as a new node and the process is repeated until the modularity no longer increases [21]. The IFS algorithm, on the other hand, simulates the flow of information to detect communities. It first identifies the influential node as the origin of information flow and then simulates information flows from this node to other nodes. All nodes that a piece of information flows through are considered to belong to the same community, and the final community detection result is obtained when the information flows through the entire network or when the algorithm reaches a steady state [12]. The TLW algorithm, meanwhile, detects community structure in temporal networks by using timestamp wandering. This method converts community detection in networks into cluster analysis in vector space, and it can also incorporate temporal information to maintain modularity at a reasonable level [22].

Table 1 Performance comparison of different community detection algorithms

| Algorithms | Louvain | IFS | TLW | This paper |
|---|---|---|---|---|
| Modularity | 0.128 | 0.153 | 0.425 | **0.524** |
| Average number of algorithms | 37.5 | 29.7 | 27.46 | **21.52** |

According to the results shown in Table 1, the modularity obtained by TLW and the algorithm proposed in this paper for the reconstructed weighted directed friendship network are higher than those obtained by the Louvain and IFS algorithms for the unweighted undirected friendship network. This implies that the reconstructed weighted directed friendship network used in this paper is better at revealing the real friendships compared to the initial unweighted undirected friendship network. Moreover, the modularity of this paper's algorithm is higher than that of the TLW algorithm. This is because the algorithm proposed in this paper can capture the exponential decay trend of friendships in continuous time and is better at detecting communities through the flow of information between friendship network. These results indicate the superiority of the algorithm proposed in this paper in capturing friendships.

### 4.3 The impact of information dissemination origins

During the process of community detection, the number of information dissemination origins can have a notable influence on the association detection results. In order to examine the effect of the proportion of information dissemination origins on the community detection results and determine the optimal proportion of information dissemination origins in the reconstructed weighted directed friendship network under continuous time, this study takes the spring semester of 2018 as an example and performs community detection with various proportions of information dissemination origins. Table 2 presents the modularity values obtained, as well as the number of communities and the average number of people per community after community detection.

As shown in Table 2, the modularity changes from small to large and then from large to small as the proportion of information dissemination origins changes from 50% to 5%. The maximum modularity of 0.552 is obtained when the proportion of information dissemination origins is 10%, indicating the best module division. However, modularity is not the only indicator of friend group division. The proportion of communities and the average number of people in a community also provide insights into the division of campus friend groups. A larger or smaller proportion of communities and average number of people in a community indicate an unreasonable division of friend groups. Therefore, in order to obtain a more reasonable division of friend groups, this paper considers the proportion of communities and the average number of people in a community in addition to modularity. After comprehensive analysis, the proportion of information dissemination origins is chosen to be 20%. At this proportion, the modularity is not the highest, but the number of communities and the average number of people in a community are more reasonable, indicating a more reasonable division of friend groups. This choice also demonstrates the effectiveness of the algorithm in capturing the underlying structure of the campus friend network.

Table 2 Community detection results with different proportion of information dissemination origins

| different proportion of information dissemination origins | Modularity | number of communities | average number of people per community |
| --- | --- | --- | --- |
| 50% | 0.313 | 1043 | 6.826 |
| 45% | 0.347 | 965 | 7.858 |
| 40% | 0.452 | 643 | 13.156 |
| 35% | 0.412 | 747 | 11.205 |
| 30% | 0.442 | 642 | 13.176 |
| 25% | 0.478 | 526 | 16.694 |
| 20% | 0.503 | 412 | 21.852 |
| 15% | 0.54 | 280 | 32.7 |
| 10% | 0.552 | 181 | 49.448 |
| 5% | 0.389 | 81 | 101.111 |

## 4.4 The evolution of campus friendship network

After reconstructing the traditional campus friendship network into a weighted directed campus friendship network in continuous time, this paper executes a community detection algorithm on it and analyzes the evolution trend of the campus friendship network over time. To demonstrate the community detection results and the changes in friendship network over different semesters, this paper calculates the number of communities and average number of people per community after community detection, with the proportion of information dissemination origins chosen as 20%. The specific results are shown in Table 3.

Table 3 Results of weighted directed network community detection for 6 semesters

| Semesters | Modularity | Number of communities | Average number of people per community |
|---|---|---|---|
| Fall Semester 2017 | 0.382 | 254 | 6.217 |
| Spring Semester 2018 | 0.422 | 304 | 8.868 |
| Fall Semester 2018 | 0.504 | 411 | 21.903 |
| Spring Semester 2019 | 0.48 | 329 | 26.812 |
| Fall Semester 2019 | 0.524 | 341 | 21.52 |
| Spring Semester 2020 | 0.368 | 591 | 25.18 |

Table 3 shows that the modularity obtained after community detection in six different semesters is all greater than 0.368, indicating the effectiveness of this paper's algorithm in dividing communities in the reconstructed weighted directed friendship network. The modularity in different semesters varies due to the total number of students. Additionally, the number of communities in different semesters and the average number of people in a community reflect changes in the campus friendship network over time to some extent. In the fall semester of 2017 and the spring semester of 2018, the amount of data was significantly smaller than that of other semesters due to an imperfect data collection system, resulting in fewer communities and smaller average community size. In the fall semester of 2018, spring semester of 2019, and fall semester of 2019, the method in this paper achieved good community detection results, as the amount of data was not much different among these semesters. The impact of COVID-19 in the fall semester of 2020 led to a significant increase in the number of students spending time on campus, resulting in the algorithm in this paper detecting more communities. However, despite differences in the number of students, the community detection algorithm in this paper identified a relatively stable average number of students per community in different semesters, with an average of about 23.85 members per community in the campus friendship network.

## 4.5 Behavior comparison of students in different communities

This paper aims to compare and analyze the behavior of students in different communities by quantitatively examining their consumption level, dining habits, and behavior regularity in four semesters. To measure the deviation of different groups of data, the variance of the data is used. Specifically, the consumption level of students is measured by their spending in the school cafeteria and supermarket, where higher spending indicates a higher consumption level. The mealtime habits are measured by the number of times and days a student spends in the school cafeteria per semester. The behavioral regularity is measured by the behavioral entropy, a metric used to measure the regularity of students' lives. In this paper, students' behavioral regularity is determined based on the entropy of their dining in the cafeteria for three meals in the morning, lunch, and evening, bathing in the bathroom, and shopping in the supermarket. Generally, the smaller the entropy, the more regular the behavior of students.

To analyze the student behavior before and after the community detection, this paper first calculates the average consumption amount, average consumption times, consumption days per semester, bathing behavior entropy, breakfast dining behavior entropy, lunch dining behavior entropy, and dinner dining behavior entropy for each semester. Then, it calculates the variance of all indicators for all students in a semester as a control, and calculates the average variance of the above indicators of students belonging to the same community after the community detection. Finally, it compares the above two variance values to analyze the student behavior before and after the community detection. For instance, to examine the bath entropy of students in the spring semester of 2019, this paper first calculates the bath entropy of each student in the spring semester of 2019 and computes the variance of the bath entropy of all students. Secondly, it carries out community detection, obtaining a total of 329 student communities, and calculates the variance of the students' bathing entropy in each community. Finally, it averages the variance of the 329 communities and compares this value with the variance of the bathing entropy of all students before the community detection.

Table 4 Analysis of student behavior before and after community detection

| Semesters | Before or after community detection | Amount | times | days | bath entropy | Breakfast entropy | Lunch entropy | Dinner entropy |
|---|---|---|---|---|---|---|---|---|
| Fall Semester 2017 | Before | 35.387 | 0.698 | 726.183 | 0.129 | 0.254 | 0.085 | 0.124 |
| | After | 27.765 | 0.568 | 596.746 | 0.104 | 0.204 | 0.067 | 0.101 |
| Spring Semester 2018 | Before | 41.477 | 0.957 | 469.297 | 0.126 | 0.337 | 0.115 | 0.163 |
| | After | 37.227 | 0.84 | 411.391 | 0.111 | 0.294 | 0.101 | 0.142 |
| Fall Semester 2018 | Before | 49.002 | 1.396 | 611.329 | 0.13 | 0.341 | 0.124 | 0.198 |
| | After | 46.711 | 1.326 | 584.806 | 0.122 | 0.324 | 0.119 | 0.189 |
| Spring Semester 2019 | Before | 58.109 | 1.534 | 845.544 | 0.141 | 0.337 | 0.117 | 0.214 |
| | After | 57.979 | 1.484 | 812.803 | 0.134 | 0.324 | 0.112 | 0.204 |

Table 4 shows that the average consumption amount, average consumption times, consumption days, bathing behavior entropy, breakfast dining behavior entropy, lunch dining behavior entropy, and dinner dining behavior entropy values of students in each semester are lower after community detection. This suggests that students belonging to the same community exhibit more similar consumption levels, dining habits, and behavioral regularity. For instance, in the spring semester of 2019, the variance of the bathing entropy of all students before community detection was 0.141, whereas after community detection, it was 0.134. A smaller variance indicates less deviation of the data and more concentration of the data. Hence, the bathing entropy of students belonging to the same community becomes more similar after community detection, indicating that students in the same community exhibit more similar bathing behavior.

The results in Table 4 also demonstrate that students in the same community exhibit more similar consumption levels, dining habits, and behavioral regularity after community detection. Additionally, since friends in the same social circle have similar characteristics to a great extent, the results in Table 4 further establish the accuracy and effectiveness of the community detection algorithm used in this paper.

## 5. Conclusion

The advancement of data-driven science and social computing has made it possible to simulate and analyze the social behavior of college students using massive data. The study of campus friendship network using large-scale objective data recorded by university campus card systems is of great research value. In this paper, we propose a new community detection algorithm for a continuous-time weighted directed friendship network to improve the effectiveness of traditional campus friendship network and the accuracy of community detection results. Our research results show that the reconstructed weighted directed friendship network can better reveal real friend relationships than the original undirected and unweighted friendship network. Moreover, the proposed community detection algorithm achieves better results. Each community in the campus friendship network has an average of approximately 23.85 members, and the campus friendship network changes over time. Especially after the COVID-19 outbreak and subsequent school closures, the number of students spending time on campus significantly increased, leading to the appearance of more communities in the campus friendship network and a significant increase in the number of students in each community. Furthermore, after community detection, the average consumption amount, consumption times, consumption days, and behavioral regularity of students are all smaller than those before community detection, indicating that after community detection, the consumption level, dining habits, and behavior regularity of students belonging to the same community are more similar. These results demonstrate the accuracy and effectiveness of the proposed community detection algorithm. This research has significant theoretical and practical implications and is highly relevant to the theoretical research of complex friendship network community detection and college student management.